\documentstyle[11pt,aaspp4,psfig]{article}
\def\rmcdm{{\rm c}}
\def\n0LC{{n^{\rm LC}_0}}
\def\rmax{{r_{\rm max}}}
\def\rmin{{r_{\rm min}}}
\def\smax{{s_{\rm max}}}
\def\smin{{s_{\rm min}}}

\def\bfx{{\bf x}}
\def\bfs{{\bf s}}
\def\bft{{\vec\gamma}}
\def\nLC{{n^{\rm LC}}}
\def\ns{{n_{\rm s}}}
\def\vLC{{\vec v^{\rm LC}}}

\def\rmax{{r_{\rm max}}}

\def\bfR{{\bf R}}
\def\bfk{{\bf k}}

\def\calS{{\cal D}}
\def\xis{{\xi_{\rm s}}}
\def\pp{\par\parshape 2 0truecm 15.5truecm 1truecm 14.5truecm\noindent}
\newcommand{\simgt}{\lower.5ex\hbox{$\; \buildrel > \over \sim \;$}}
\newcommand{\simlt}{\lower.5ex\hbox{$\; \buildrel < \over \sim \;$}}

\begin{document}

\parbox{\hsize}{
\begin{flushright}
HUPD-9830 ~~~~~~~~~~~\\
\vspace{-1mm}
December 1998 ~~~~~~~~~~~\\
\vspace{-1mm}
Revised March 1999 ~~~~~~~~~~~\\
\end{flushright}}

\title{Two-point correlation function of high-redshift objects 
on a light-cone~: \\
Effect of the linear redshift-space distortion}

\bigskip
\author{Hiroaki Nishioka and Kazuhiro Yamamoto}
\affil{ Department of Physics, Hiroshima
    University, Higashi-Hiroshima 739-8526, Japan} 

\bigskip


\bigskip

\received{1998 December}
\accepted{}

\begin{abstract}

A theoretical formulation for the two-point correlation function
on a light-cone is developed in the redshift space. On the basis 
of the previous work by Yamamoto \& Suto (1999), in which a 
theoretical formula for the two-point correlation function on a 
light-cone has been developed in the real space, we extend it to 
the formula in the redshift space by taking the peculiar velocity 
of the sources into account. A simple expression for the two-point 
correlation function is derived. We briefly discuss QSO correlation 
functions on a light-cone adopting a simple model of the sources.

\end{abstract}


\clearpage

\section{Introduction}

The clustering of high-redshift objects is one of the current topics
in the fields of observational cosmology and astrophysics.
The high-redshift objects of $z\simgt1$ are becoming fairly common, and
evidences of the clustering nature of such cosmic objects 
are reported in the various observational bands,
e.g., X-ray selected AGNs (Carrera et~al. 1998), the FIRST survey 
(Cress et~al. 1996; Magliocchetti et~al. 1998), high-redshift galaxies 
(Steidel et~al. 1998; Giavalisco et~al. 1998),
and QSO surveys (Croom \& Shanks 1996; Boyle et al. 1998).
The statistics of these high-redshift objects are becoming 
higher, and we will be able to discuss the clustering 
at a quantitative level precisely in near future.
From a theoretical point of view, the most important subject
is to clarify the physical process of the formation history of
these objects. The standard theoretical framework for the cosmic 
structure formation is based on the cold dark matter (CDM) model 
with gaussian initial density fluctuations.
The clustering nature of the high-redshift objects
provides us with many kinds of tests for the theoretical models 
(e.g., Peacock 1998; Jing \& Suto 1998). 

When analyzing the clustering nature of the high-redshift objects
at a quantitative level, we must take the light-cone effect
into account properly. Namely, such cosmological observations 
are feasible only on the light-cone hypersurface defined 
by the current observer. And the effect of the time-evolution
of the sources, i.e., the luminosity function, 
the clustering amplitude, and the bias, contaminates an 
observational data.
Thus this light-cone effect is especially important to discuss the 
three-dimensional two-point correlation function of the 
high-redshift objects. Some aspects of the light-cone effect has been 
discussed (Matarrese et~al. 1997;Matsubara, Suto, \& Szapudi 1997; 
Nakamura, Matsubara, \& Suto 1998;Laix \& Starkman 1998).
Recently one of the authors (K.Y.) \& Suto developed a formulation
for the two-point correlation function for the high-redshift 
objects defined on the light-cone hypersurface 
(Yamamoto \& Suto 1999: hereafter Paper~I).
The expression for the two-point correlation function on the light-cone 
was derived in a rigorous manner starting from first principle 
corresponding to the conventional pair-count analysis.
This investigation is very important because it gives a rigorous
relation between an observational data processing and a theoretical
prediction as to the two-point correlation function on a light-cone 
for the first time. However this investigation is restricted 
to the formula in the real space, though observational maps
of the high-redshift objects are obtained in the redshift space.

It is well known that the peculiar velocity of sources
distorts their distribution in the redshift space
(e.g., Davis \& Peebles 1983; Kaiser 1987; Hamilton 1997).
And this effect has been discussed as a probe of cosmological
density parameters (e.g., Szalay, Matsubara, \& Landy 1998; Nakamura, 
Matsubara, \& Suto 1998; Matsubara \& Suto 1996; 
Hamilton \& Culhane 1996; Heavens \& Taylor 1995; Suto, et~al. 1999).
In the previous paper (Paper~I) the effect of the redshift-space distortion 
due to the peculiar velocity of the sources is not taken into 
account because it was formulated in the real space.
From a practical point of view, the formula  
in the redshift space must be developed.
The purpose of the present paper is to develop such a theoretical 
formula for the two-point correlation function on a light-cone
hypersurface by taking the redshift-space distortion due to the peculiar 
velocity into account. 

The paper is organized as follows: In \S 2 we develop 
a formulation for the two-point correlation function 
on the light-cone hypersurface in the redshift space
in order to incorporate the linear redshift-space distortion. 
The expression for the two-point correlation function
is presented in a rather simple form by using appropriate
approximations. The main result is equation (\ref{WRB}).
As a demonstration of the usefulness of our
formalism, we apply the formula to QSO correlation functions 
adopting a simple model of source distribution and cosmological models. 
The validity of the plane-parallel, or distant observer, 
approximation, for the correlation function of high-redshift objects
is also discussed. \S 4 is devoted to discussion and conclusion.
Throughout this paper we use the units in which 
the light velocity $c$ is unity.

\section{Two-point correlation function in the redshift space}
In this section we develop a theoretical formulation for the 
two-point correlation function on a light-cone hypersurface 
in the redshift space by taking the peculiar motion of sources 
into account. In the present paper, we focus on the 
spatially-flat Friedmann-Lemaitre universe, 
whose line element is expressed in terms of the conformal time $\eta$ as
\begin{equation}
  ds^2 = a^2(\eta) \left[-d\eta^2+d\chi^2+\chi^2 d\Omega_{(2)}^2 \right] .
\label{metric}
\end{equation}
Here the scale factor is normalized to be unity at present, i.e., 
$a(\eta_0)=1$. The Friedmann equation is 
\begin{equation}
  \biggl({\dot a\over a}\biggr)^2=
  H_0^2\biggl({\Omega_0\over a}+a^2 \Omega_\Lambda\biggr),
\label{Friedman}
\end{equation}
where $\Omega_\Lambda=1-\Omega_0$, the dot denotes $\eta$ differentiation,
and $H_0$ is the Hubble constant $H_0=100h$ km/s/Mpc.

Since our fiducial observer is located at the origin of the
coordinates ($\eta=\eta_0$, $\chi=0$), an object at $\chi$ and 
$\eta$ on the light-cone hypersurface of the observer 
satisfies a simple relation $\eta=\eta_0-\chi$. 
Then the (real-space) position of the source on the 
light-cone hypersurface is specified by $(\chi,\bft)$,
where $\bft$ is an unit directional vector.
In order to avoid confusion, we introduce
the radial coordinate $r$ instead of $\chi$,  
and we denote the metric of the three-dimensional real space on which
the observable sources are distributed, as follows,
\begin{equation}
  ds_{LC}^2 = dr^2+r^2d\Omega_{(2)}^2.
\label{LCmetric}
\end{equation}
Denoting the comoving number density of observed objects at a conformal
time $\eta$ and at a position $(\chi,\bft)$ by 
$n(\eta,\chi,\bft)$, then the corresponding number
density projected onto the space (\ref{LCmetric}) is obtained by
\begin{equation}
  \nLC(r,\bft)=n(\eta,\chi,\bft)~\bigr|_{
  \eta\rightarrow \eta_0-r, ~\chi\rightarrow r}~.
\label{eq:nlc1}
\end{equation}
Introducing the mean {\it observed} (comoving) number density
$n_0(\eta)$ at time $\eta$ and the density fluctuation of 
luminous objects $\Delta(\eta,\chi,\bft)$, we write
\begin{equation}
  n(\eta,\chi,\bft) = n_0(\eta) \left[1+\Delta(\eta,\chi,\bft)\right],
\end{equation}
then equation (\ref{eq:nlc1}) is rewritten as
\begin{equation}
  \nLC(r,\bft)=n_0(\eta) \left[1+\Delta(\eta,\chi,\bft) 
  \right]~\bigr|_{\eta\rightarrow \eta_0-r, ~\chi\rightarrow r} .
\label{nLC}
\end{equation}
Note that the mean {\it observed} number density $n_0(\eta)$ is
different from the mean number density of the objects
$\overline{n}(\eta)$ at $\eta$ by a factor of the selection function
$\phi(\eta)$ which depends on the luminosity function of the objects
and thus the magnitude-limit of the survey, for instance:
$
  n_0(\eta) = \overline{n}(\eta) \phi(\eta).
$

In the similar way, if we know the peculiar velocity field, 
the corresponding quantity projected
onto the space (\ref{LCmetric}) is obtained by
\begin{equation}
  \vLC(r,\bft)={\vec v}_{\rmcdm}(\eta,\chi,\bft)~\bigr|_{
  \eta\rightarrow \eta_0-r, ~\chi\rightarrow r}~,
\label{eq:vlc1}
\end{equation}
where ${\vec v}_{\rmcdm}(\eta,\chi,\bft)$ is the CDM velocity field.
Here we assume that the peculiar velocity field of luminous
objects agrees with the CDM velocity field.

In Appendix A we summarized equations for the linear perturbation
theory in the CDM dominated universe. Thus the linearized 
CDM density perturbation can be solved completely. However, 
the evolution of the source density fluctuations can not 
be solved completely since the bias mechanism is not well
understood at present unfortunately. 
Then we must assume a model for the bias which connects 
the CDM density perturbations and the source number density 
fluctuations. In the present paper we assume the scale-dependent
bias model:
\begin{equation}
  b(k;\eta)={\Delta_{klm}(\eta)\over \delta^{\rm (c)}_{klm}(\eta)},
\label{ddb}
\end{equation}
where $\Delta_{klm}(\eta)$ and $\delta^{\rm (c)}_{klm}(\eta)$
are the Fourier coefficients for the source number density fluctuation
and the CDM density fluctuation, respectively (see also Appendix A).

The next task is to describe the relation between the real space 
and the redshift space, since we consider the distribution 
of sources in the redshift space.
First we consider how the peculiar velocity of a source
distorts the estimation of the distance to the source. 
Let us assume that a source at redshift $z$ (at a position $(r,\bft)$ 
in the real space) is moving with a peculiar velocity $\vec v$. 
The observed photon frequency
$\nu_{\rm obs}$ and the emitted photon frequency
$\nu_{\rm emit}$ is related as 
\begin{equation}
  \nu_{\rm obs}={\nu_{\rm emit}\over 1+z}(1-{\vec \gamma}\cdot{\vec v}).
\end{equation}
From this equation, we find the shift in the apparent redshift due to the 
peculiar velocity as
\begin{equation}
  \delta z=(1+z)({\vec \gamma}\cdot{\vec v})~.
\label{deltaz}
\end{equation}
{}From the Friedmann equation (\ref{Friedman}), we have
\begin{equation}
   \delta \eta=-{1\over H_0}{a^{3/2}\delta z\over 
  \sqrt{\Omega_0+\Omega_\Lambda a^3}}~.
\label{deltaeta}
\end{equation}
Combining equations (\ref{deltaz}) and (\ref{deltaeta}), then
we obtain the apparent shift in the comoving coordinate due to 
the peculiar velocity as 
\begin{equation}
   \delta r= -\delta \eta = 
 {{\cal Z}(\eta) \over H_0} ~\vec \gamma\cdot\vec v~\bigr|_{
 \eta\rightarrow \eta_0-r, ~\chi\rightarrow r}~,
\label{deltar}
\end{equation}
where we defined
\begin{equation}
  {\cal Z}(\eta) =
   {a(\eta)^{1/2}\over \sqrt{\Omega_0+\Omega_\Lambda a(\eta)^3}} ~.
\end{equation}

We introduce the variable $s$ to denote the radial coordinate 
in the redshift space.
Then a position in the redshift space is specified by $(s,\bft)$,
while the real space is done by $(r,\bft)$.
The relation between the redshift position
and the real position is
\begin{equation}
  s=r+\delta r,
\label{srr}
\end{equation}
where $\delta r$ is specified by equation (\ref{deltar}).
The conservation of the number of sources gives (Hamilton 1997)
\begin{equation}
  \ns(s,\bft)s^2dsd\Omega_{\bft}=n^{\rm LC}(r,\bft)r^2drd\Omega_{\bft},
\label{nsnLC}
\end{equation}
where $\ns(s,\bft)$ denotes the number density in the redshift space
and $n^{\rm LC}(r,\bft)$ does the number density in the real space.
These two equations (\ref{srr}) and (\ref{nsnLC}) specify
the relation between the redshift space and the real space.

Now let us consider the two-point correlation function in the 
redshift space. We start from the following ensemble estimator 
for the two-point correlation function:
\begin{equation}
  {\cal X}_{\rm s}(R)={1\over V^{\rm LC}}
  \int{d\Omega_{\hat \bfR}\over 4\pi} 
  \int ds_1 s_1^2 d\Omega_{\bft_1} 
  \int ds_2 s_2^2 d\Omega_{\bft_2}
  \ns(s_1,\bft_1)\ns(s_2,\bft_2)
  \delta^{(3)}(\bfs_1-\bfs_2-\bfR),
\label{C3}
\end{equation}
where $\bfs_1=(s_1,\bft_1)$ and $\bfs_2=(s_2,\bft_2)$ and
$R=|\bfR|$, $\hat \bfR=\bfR/R$, and $V^{\rm LC}$ is the comoving
survey volume of the data catalogue:
\begin{equation}
  V^{\rm LC} =\int_\smin^\smax  s^2 ds 
    \int d\Omega_\bft={4\pi \over 3} (\smax^3 - \smin^3) ,
\end{equation}
with $\smax = s(z_{\rm max})$ and $\smin= s(z_{\rm min})$ being the
boundaries of the survey volume. Equation (\ref{C3}) is a natural 
extension of the ensemble estimator for the two-point correlation 
function in the redshift space (see also Paper~I).

By using equations (\ref{srr}) and (\ref{nsnLC}), 
we rewrite equation (\ref{C3}) 
in terms of the variables in the real space:
\begin{eqnarray}
  &&{\cal X}_{\rm s}(R)={1\over V^{\rm LC}}
  \int{d\Omega_{\hat \bfR}\over 4\pi} 
  \int dr_1 r_1^2 d\Omega_{\bft_1} 
  \int dr_2 r_2^2 d\Omega_{\bft_2}
  \nLC(r_1,\bft_1)\nLC(r_2,\bft_2)
\nonumber
\\
  &&\hspace{5cm}
  \times\delta^{(3)}(\bfx_1+\delta\bfx_1-\bfx_2-\delta\bfx_2-\bfR),
\label{C4}
\end{eqnarray}
where $\bfx_1+\delta\bfx_1=(r_1+\delta r_1,\bft_1)$, 
      $\bfx_2+\delta\bfx_2=(r_2+\delta r_2,\bft_2)$, 
and $\delta r_1$ and $\delta r_2$ are given by (\ref{deltar}).
Then we approximate as 
\begin{equation}
  \delta^{(3)}(\bfx_1+\delta\bfx_1-\bfx_2-\delta\bfx_2-\bfR)\simeq
  \biggl(1+\delta\bfx_1\cdot{\partial\over \partial \bfx_1}\biggr)
  \biggl(1+\delta\bfx_2\cdot{\partial\over \partial \bfx_2}\biggr)
\delta^{(3)}(\bfx_1-\bfx_2-\bfR),
\end{equation}
where we can write $\delta\bfx\cdot{\partial\big/\partial\bfx}
=\delta r{\partial\big/\partial r}$ since only the radial component
of $\delta\bfx$ has a non-zero value.
By using this approximation and equation (\ref{nLC}) 
we derive the following equation from (\ref{C4}): 
\begin{eqnarray}
  &&{\cal X}_{\rm s}(R)={1\over V^{\rm LC}}
  \int{d\Omega_{\hat \bfR}\over 4\pi} 
  \int dr_1 r_1^2 d\Omega_{\bft_1} 
  \int dr_2 r_2^2 d\Omega_{\bft_2}
  \n0LC (r_1) \n0LC (r_2)
\nonumber
\\
  &&\hspace{1.5cm}
  \times\prod_{i=1}^{2}
  \biggl[1+\Delta(r_i,\bft_i)+\delta r_i {\partial\over \partial r_i}
  \biggr]\delta^{(3)}(\bfx_1-\bfx_2-\bfR),
\label{C6}
\end{eqnarray}
where we used the notations:
\begin{eqnarray}
  \n0LC(r)=n_0(\eta)~\bigr|_{\eta\rightarrow \eta_0-r} ~,
\hspace{1cm}
  \Delta(r,\bft)=\Delta(\eta,\chi,\bft) 
  ~\bigr|_{\eta\rightarrow \eta_0-r, ~\chi\rightarrow r} ~,
\end{eqnarray}
and $\delta r_i$ is understood as
\begin{equation}
   \delta r_i= {{\cal Z}(\eta)\over H_0} ~\vec \gamma\cdot\vec v~\bigr|_{
\eta\rightarrow \eta_0-r_i, ~\chi\rightarrow r_i}~,
\label{delta}
\end{equation}
where $i=1,2$.

Next we consider the ensemble average of the ensemble estimator 
${\cal X}_{\rm s}(R)$.
Since $\delta r_1$ and $\delta r_2$ are the order of linear perturbation,
then the ensemble average is written as
\begin{equation}
\bigl<{\cal X}_{\rm s}(R)\bigr>={\cal U}(R)+{\cal W}_{\rm s}(R),
\end{equation}
where
\begin{eqnarray}
  &&{\cal U}(R)=
  {1\over V^{\rm LC}}
  \int{d\Omega_{\hat \bfR}\over 4\pi} 
  \int dr_1 r_1^2 \int d\Omega_{\bft_1} 
  \int dr_2 r_2^2 \int d\Omega_{\bft_2}
  \n0LC (r_1)\n0LC (r_2)
  \delta^{(3)}(\bfx_1-\bfx_2-\bfR),
\nonumber
\\
\label{C33}
\end{eqnarray}
and
\begin{eqnarray}
  &&{\cal W}_{\rm s}(R)=
  {1\over V^{\rm LC}}
  \int{d\Omega_{\hat \bfR}\over 4\pi} 
  \int dr_1 r_1^2 \int d\Omega_{\bft_1} 
  \int dr_2 r_2^2 \int d\Omega_{\bft_2}
  \n0LC(r_1) \n0LC(r_2)
\nonumber
\\
  &&
 \hspace{1cm}\times
  \biggl<\biggl(\Delta(r_1,\bft_1)+\delta r_1 {\partial\over \partial r_1}
  \biggr)
  \biggl(\Delta(r_2,\bft_2)+\delta r_2 {\partial\over \partial r_2}
  \biggr)\biggr>
  \delta^{(3)}(\bfx_1-\bfx_2-\bfR).
\label{C7}
\end{eqnarray}

In Appendix B we presented the explicit calculations 
for ${\cal W}_{\rm s}(R)$. According to the result, equation 
(\ref{C7}) reduces to the following form within the linear 
theory of perturbation:
\begin{eqnarray}
  &&{\cal W}_{\rm s}(R)=
  {1\over V^{\rm LC}}{1\over \pi R}
  \int\int_{\cal S} dr_1 dr_2 r_1 r_2 
  \n0LC(r_1) \n0LC(r_2) D_1(\eta_0-r_1)D_1(\eta_0-r_2)  
\nonumber
\\
  &&
 \hspace{1cm}\times  \int dk k^2 P(k)
  \prod_{i=1}^{2}
  \Bigl[b(k;\eta_0-r_i) - k^{-2}\calS_{r_{i}}\Bigr]
j_0\Bigl(k\sqrt{r_1^2+r_2^2-2r_1r_2\cos\theta}\Bigr)~,
\label{C68}
\end{eqnarray}
where 
\begin{equation}
  \calS_r=f(\eta_0-r){\partial^2\over \partial r^2}+f(\eta_0-r){d\over dr}
   \ln\Bigl[r^2 n_0^{\rm LC}(r)D_1(\eta_0-r)f(\eta_0-r)\Bigr]
  {\partial\over \partial r}~,
\label{defcalS}
\end{equation}
and $P(k)$ is the CDM power spectrum at present, 
$D_1(\eta)$ is the linear growth rate normalized to be unity at 
present, $f(\eta)$ is defined as $f(\eta)=d \ln D_1(\eta)/d \ln a(\eta)$, 
and $\cos\theta$ should be replaced by 
$\cos\theta=(r_1^2+r_2^2-R^2)/ 2r_1r_2$ after operating
the differentiations with respect to $r_1$ and $r_2$. 

Omitting the second term in the derivative (\ref{defcalS}),
equation (\ref{C68}) reduces to the simple form 
in the case $R\ll \rmin$ and $R\ll \rmax$ (see Appendix B),
\begin{eqnarray}
  &&{\cal W}_{\rm s}(R)\simeq
  {4\pi\over V^{\rm LC }}
  \int_\rmin^\rmax dr r^2 \n0LC(r)^2 
  {1\over 2\pi^2}\int  k^2 dk  P(k)  
  \Bigl[b(k;\eta_0-r)D_1(\eta_0-r)\Bigr]^2
\nonumber
\\
  && \hspace{1cm}\times
  \biggl[1+{2\over 3}\beta(\eta_0-r)+{1\over5}\beta(\eta_0-r)^2
  \biggr]j_0(kR)~,
\label{WRFinal}
\end{eqnarray}
where $\beta(k;\eta)$ is defined by 
\begin{equation}
  \beta(k;\eta)={f(\eta)\over b(k;\eta)}=
  {1\over b(k;\eta)}{d\ln D_1(\eta)\over d\ln a(\eta)}~,
\label{defbeta}
\end{equation}
and we assumed $\smax=\rmax~(\smin=\rmin)$.

We can derive the following equation from a similar calculation
in the above (see also Paper~I):
\begin{eqnarray}
  {\cal U}(R)&\simeq&
  {4\pi\over V^{\rm LC }} \int_\rmin^{\rmax}   r^2 dr
   \n0LC(r)^2 .
  \label{C81}
\end{eqnarray}
Following Paper~I, we define the two-point correlation 
function on the light-cone hypersurface:
\begin{equation}
  \xis^{\rm LC}(R) ={\bigl<{\cal X}_{\rm s}(R)^{}\bigr>-{\cal U}(R)
  \over {\cal U}(R)}={{\cal W}_{\rm s}(R)\over {\cal U}(R)}~. 
\label{defxiA} 
\end{equation}
Substituting equations (\ref{WRFinal}) and (\ref{C81}) into 
(\ref{defxiA}), we have 
\begin{eqnarray}
  &&\xis^{\rm LC}(R)\simeq
  \biggl[\int_\rmin^{\rmax}    dr r^2
   \n0LC(r)^2 \biggr]^{-1}
  \int_\rmin^\rmax dr r^2 \n0LC(r)^2 
\nonumber
\\
 &&\hspace{1.5cm}\times
  {1\over 2\pi^2}\int  k^2 dk  P(k)  
  b(k;\eta_0-r)^2D_1(\eta_0-r)^2 
\nonumber
\\
 &&\hspace{1.5cm}\times
  \biggl[1+{2\over 3}\beta(\eta_0-r)+{1\over5}\beta(\eta_0-r)^2
  \biggr]j_0(kR)~.
\label{WRB}
\end{eqnarray}
This is the final expression for the two-point correlation 
function on the light-cone hypersurface in which the linear 
redshift-space distortion is taken into account. 
Comparing this result with $\xi_A^{\rm LC}(R)$ in Paper~I, 
the terms in proportion to $\beta(k;\eta)$ 
are the new terms which represent the effect of the 
linear redshift-space distortion.

\section{A Simple Demonstration}
In this section we apply the formula developed in the previous
section to QSO two-point correlation function. Evidence for the spatial
correlation in the QSO-distribution is reported (Croom \& Shanks 1996;
Boyle et~al. 1998), however, it seems difficult to draw 
definite cosmological conclusions from the comparison 
with the currently available data. Then we only demonstrate 
the usefulness of our formalism by calculating the QSO two-point 
correlation function based on a simplified model for the distribution 
and the bias model. As for the bias, we here consider the 
scale-independent bias model by Fry (1996):
\begin{equation} 
  b(\eta)= 1 +{1\over D_1(\eta)}  (b_0-1),
\end{equation}
where $b_0$ is a constant parameter. Note that the bias $b(\eta)$
at high-redshift becomes larger as $b_0$ becomes larger.
Here we also assume that the sources are distributed in the range 
$0.3\leq z\leq 3$ with a constant number 
density, i.e.,  $n_0={\rm const}$. This model may be 
over-simplified, however, we have checked that the 
qualitative features have not been changed even when
adopting more realistic models in Paper~I.

In figures~1 and 2 we show the two-point correlation function 
$\xis^{\rm LC}(R)$ and other mass correlation functions for comparison.
We show the case for the standard cold dark matter (SCDM) model in Fig.~1, 
in which we adopted $\Omega_0=1,~\Omega_\Lambda=0,~h=0.5,~$ and the CDM 
density power spectrum normalized as $\sigma_8=0.56$
(Kitayama \& Suto 1997). The case for the cosmological model
with a cosmological constant ($\Lambda$CDM model) 
is shown in Fig.2, in which  $\Omega_0=0.3,~\Omega_\Lambda=0.7,
~h=0.7, ~\sigma_8=1.0$ are adopted.
In each panel (a)-(c), the three lines show the correlation functions
on a light-cone.
The solid line shows $\xis^{\rm LC}(R)$ of equation (\ref{WRB}). 
The dashed line shows the case when neglecting the effect of 
the redshift-space distortion by setting $\beta=0$ in (\ref{WRB}).
On the other hand, the panel (d) shows the linear and nonlinear 
mass two-point correlation functions defined on a constant time 
hypersurface $z=0$, $1$, and $2$.

From these figures it is apparent that the larger bias at the
high-redshift derives the larger amplitude of the correlation function
on a light-cone. Furthermore the effect of the redshift-space 
distortion always 
amplifies the correlation function from comparing the solid line and the 
dashed line in the panels (a)-(c), as expected. However the relative 
difference between the solid line and the dashed line becomes smaller
as the bias becomes large and more effective. 
This is an expected feature because $\beta$-factor
becomes smaller as the bias becomes larger 
(see eq.[\ref{defbeta}]).

We have also calculated the correlation function by adopting the exact 
expression (B20) instead of (28). The difference is less than $1~\%$
for $R\simlt 100 h^{-1}{\rm Mpc}$, and is negligible. Thus the formula 
(32) is a well approximated formula, and is an useful expession for
the correlation function for high-redshift objects. The expression (32) 
is easily understood in an intuitive manner. Namely, the linear power 
spectrum in the redshift space is amplified by $(1+\beta\mu_\bfk)^2$ 
over its unredshifted counterpart $P(k)$ in the plane-parallel 
approximation, where $\beta$ is defined in the 
same way as (\ref{defbeta}) and $\mu_\bfk=\vec\gamma\cdot\bfk/k$ 
(e.g., Kaiser 1987; Hamilton 1997). 
This formula leads that the angle-averaged redshift power
spectrum is amplified by the factor, $(1+2\beta/3+\beta^2/5)$, 
over the unredshifted power spectrum.
Thus (32) is the expected formula obtained by multiplying the factor, 
$(1+2\beta/3+\beta^2/5)$, at each cosmological time
over unredshifted counterpart $P(k)$
in the correlation function in the real space.
In this sense the formula (32) is based on the plane-parallel 
approximation.\footnote{We thank T.~Matsubara for his comment 
that our formula reproduces the formula inspired from such a
consideration based on the plane-parallel approximation.}
And our investigation shows that the use of the plane-parallel 
approximation is valid for the (angle-averaged) correlation 
function of high-redshift objects on a light-cone.

As we plotted absolute values of the two-point correlation 
functions in Figs.~1 and 2, then we can regard that the 
solid line and the dashed line show the anti-correlation 
at the large separation $R\simgt$ a few $\times 10h^{-1}{\rm Mpc}$
in the SCDM model (see Fig.~1).
Equation (32) implies that the zero-point of the correlation function 
is invariant even when the redshift space distortion is taken into 
account, as long as the bias does not depend on the scale $k$.
The critical correlation length, where the correlation changes
to the anti-correlation, is given by
\begin{equation}
  {R}={16.6 h^{-1}{\rm Mpc} \over \Omega_0 h 
  \exp[-\Omega_b-\sqrt{2h}\Omega_0/\Omega_b]},
\end{equation}
where we assumed the Harrison-Zeldovich initial density power 
spectrum and used the fitting formula for the transfer function 
(Bardeen ~et.al 1986; Sugiyama 1995). 
This critical correlation length may be observed in the upcoming 2dF 
and SDSS QSO surveys, and may be tested for the cosmological models 
and the theoretical models of bias.

\section{Summary and Discussion}

In this paper we have developed a theoretical formulation for the 
two-point correlation function for high-redshift objects on a light-cone 
in the redshift space. Our formula has been developed by extending 
the previous work (Paper~I) to the formula in the redshift space. 
We have started our formulation from considering the ensemble 
estimator of the two-point correlation function in the redshift space, 
then we have calculated the ensemble average of the estimator. 
Thus our formula has been derived in a rigorous manner starting 
from first principle corresponding to the conventional pair-count analysis. 
The calculation was cumbersome, however, a rather simple expression 
(\ref{WRB}) has been derived.

We have demonstrated the effect of the redshift-space distortion
by showing the QSO two-point correlation function adopting a
very simple model of the source distribution and the bias,
though it seems premature to draw definite cosmological 
conclusions from comparison with currently available data.
As discussed in the below, our model adopted in this paper may
be oversimplified in order to compare with a real data sample. 
Nevertheless our investigation is instructive and we 
have shown how the redshift-space distortion affects 
the correlation function for high-redshift objects on a light-cone 
(section 3). 
Our investigation shows that the redshift-space distortion becomes 
a small effect for time-varying bias models which have large 
values at high-redshift. 
The validity of the plane-parallel approximation is also shown
for the correlation function of high-redshift objects on a light-cone.

There remain uncertainties to make precise theoretical predictions. 
First we did not attempt to examine possible bias models other than 
the model by Fry (1996). However, theoretical investigations for the 
time and scale dependent bias are just beginning 
(Fry 1996; Mo \& White 1996; Dekel \& Lahav 1998; Tegmark \& Peebles 
1998; Taruya, Koyama \& Soda 1998). 
Conversely, the clustering of the high-redshift objects will be a good 
tool to test the bias models. Second we did not consider the 
realistic model for the time-evolution of number density in calculating
QSO two-point correlation function. As for this point finite 
solution will be obtained in upcoming 2dF and SDSS QSO surveys.
Third we have only considered the linear theory of the density 
perturbations, and the nonlinear effect was not considered here.
According to the previous work (Paper~I), the non-linearity of 
the source density fluctuation becomes important only at small 
separation $R\simlt$ a few $h^{-1}{\rm Mpc}$ (see also panel (d) 
in Figs. 1 and 2). And the effect of the non-linearity seems to be 
negligible at the large separation. However, the nonlinear effect in the
redshift space has not been well understood especially for high-redshift 
objects, it must be investigated in future work.
Probably numerical approaches would be needed for that purpose.


\bigskip
\bigskip
\begin{center}
{\bf ACKNOWLEDGMENTS}
\end{center}
We thank Y.~Kojima, Y.~Suto and T.~T.~Nakamura for useful discussions and 
comments. We thank T.~Matsubara for his crucial comments on the earlier 
manuscript. This research was supported in part by the Grants-in-Aid by 
the Ministry of Education, Science, Sports and Culture of Japan (09740203).

\newpage
\begin{appendix}
\begin{center}
{\bf APPENDIX}
\end{center}
\section{Review of the linear theory of the CDM density perturbations}
In this Appendix we summarize equations for the linear theory 
of the CDM density perturbations and explain the notations which 
are used in the present paper.
The linearized CDM density perturbation in the CDM dominated universe 
obeys the following equations:
\begin{eqnarray}
  && \dot \delta_{\rmcdm}+ v_{\rmcdm}^{i}{}_{|i}=0,
\label{contieq}
\\
  && \dot v_{\rmcdm}^{i}+ {\dot a\over a}v_{\rmcdm}^i+ \Psi^{|i}=0,
\label{eulereq}
\\
  && \Psi^{|i}_{\ ~|i} 
  =4\pi G \rho a^2 \delta_{\rmcdm}
  ={3 \Omega_0 H_0^2\over 2 a} \delta_{\rmcdm},
\label{poisseq}
\end{eqnarray}
where $\delta_{\rmcdm}$ is the CDM density contrast,
$v_{\rmcdm}^i$ is the CDM velocity field, 
$\Psi$ is the gravitational potential, which 
follows the gravitational poisson equation (\ref{poisseq}),
and $|i$ denotes the covariant derivative on the three-dimensional 
space.

As we are interested in the scalar perturbation, we expand the 
CDM density contrast $\delta_{\rmcdm}$ and the velocity field 
$v_{\rmcdm}^i$ in terms of the scalar harmonics as follows
(e.g., Kodama \& Sasaki 1984):
\begin{eqnarray}
  &&\delta_{\rmcdm}(\eta,\chi,\bft)
  =\int_0^\infty dk \sum_{l,m} \delta^{\rm (c)}_{klm}(\eta)
    {\cal Y}_{klm}(\chi,\bft),
\label{deltacdm}
\\
  &&v_{\rmcdm}^i(\eta,\chi,\bft)
  =\int_0^\infty dk \sum_{l,m} v_{klm}(\eta)
    {\cal Y}_{klm}^i(\chi,\bft),
 \label{C2}
\end{eqnarray}
where ${\cal Y}_{klm}$ is the normalized scalar harmonics:
\begin{eqnarray}
  &&{\cal Y}_{klm}(\chi,\bft)= X_{k}^{l}(\chi) Y_{lm}(\Omega_{\bft}),
\end{eqnarray}
with 
\begin{equation}
   X_{k}^{l}(\chi)=\sqrt{{2\over \pi}} k j_l(k\chi),
\label{Xkl}
\end{equation}
$Y_{lm}(\Omega_{\bft})$ and $j_l(x)$ are the spherical harmonics and
the spherical Bessel function, respectively, $k$ denotes
the eigenvalue of the eigen-equation:
${\cal Y}_{klm}{}^{|i}{}_{|i}=-k^2{\cal Y}_{klm}$, and 
${\cal Y}_{klm}^i$ is defined as
\begin{eqnarray}
  &&{\cal Y}_{klm}^i(\chi,\bft)=-{1\over k}{\cal Y}_{klm}(\chi,\bft){}^{|i}.
\end{eqnarray}
{}From the linearized perturbation equations (\ref{contieq})$-$(\ref{poisseq}),
we have
\begin{eqnarray}
  && \dot \delta^{\rm (c)}_{klm}+ kv_{klm}=0,
\label{contieqk}
\\
  && \dot v_{klm} + {\dot a\over a}v_{klm}-k \Psi_{klm}=0,
\\
  &&k^2\Psi_{klm}=-{3 \Omega_0 H_0^2\over 2 a} \delta^{\rm (c)}_{klm},
\end{eqnarray}
where $\Psi_{klm}$ is the Fourier coefficient defined 
in the same way as (\ref{deltacdm}).
Combining these equations, we have
\begin{equation}
  \ddot\delta^{\rm (c)}_{klm}+{\dot a\over a}\dot\delta^{\rm (c)}_{klm}
  -{3\over2}{\Omega_0H_0^2\over a}\delta^{\rm (c)}_{klm}=0.
\end{equation}
In the Friedmann-Lemaitre universe, the growing mode solution is well known:
\begin{equation}
  \delta^{\rm (c)}_{klm}(\eta)=\delta^{\rm (c)}_{klm}(\eta_0)D_1(a),
\label{ddD}
\end{equation}
with
\begin{equation}
  D_1(a)=A\sqrt{{\Omega_0\over a^3}+1-\Omega_0}
  \int_0^a da'\biggl({a'\over \Omega_0+a'^3(1-\Omega_0)}\biggr)^{3/2}.
\end{equation}
Here $A$ is a constant to be determined so that $D_1$ 
is unity at present.
From equations (\ref{contieqk}) and (\ref{C2}), we finally have
\begin{equation}
  v^i_{\rmcdm}(\eta,\chi,\bft)=\int_0^\infty dk \sum_{l,m} 
  {\dot\delta^{\rm (c)}_{klm}(\eta) \over k^2}
  {\cal Y}_{klm}(\chi,\bft){}^{|i}.
\label{vfinal}
\end{equation}

\section{Calculation of ${\cal W}_{\rm s}(R)$}
In this Appendix we present an explicit calculation of ${\cal W}_{\rm s}(R)$:
\begin{eqnarray}
  &&{\cal W}_{\rm s}(R)=
  {1\over V^{\rm LC}}
  \int{d\Omega_{\hat \bfR}\over 4\pi} 
  \int dr_1 r_1^2 \int d\Omega_{\bft_1} 
  \int dr_2 r_2^2 \int d\Omega_{\bft_2}
  \n0LC(r_1) \n0LC(r_2)
\nonumber
\\
  &&
 \hspace{1cm}\times
  \biggl<\biggl(\Delta(r_1,\bft_1)+\delta r_1 {\partial\over\partial r_1}
  \biggr)
  \biggl(\Delta(r_2,\bft_2)+\delta r_2 {\partial\over\partial r_2}
  \biggr)\biggr>
  \delta^{(3)}(\bfx_1-\bfx_2-\bfR).
\label{AC7}
\end{eqnarray}
Here $\Delta(r,\bft)$ and $\delta r(r,\bft)$ are explicitly written 
as
\begin{eqnarray}
  &&\Delta(r,\bft)
  =\int_0^\infty dk \sum_{l,m} 
  \delta^{\rm (c)}_{klm}(\eta_0)b(k;\eta_0-r)D_1(\eta_0-r)
  {\cal Y}_{klm}(r,\bft),
\label{expA}
\\
  &&\delta r(r,\bft)={{\cal Z}(\eta_0-r)\over H_0}
  v_{\rmcdm}^r(\eta_0-r,r,\bft)
\nonumber
\\
  &&\hspace{1.4cm}
  =
  \int_0^\infty dk \sum_{l,m} 
  \delta^{\rm (c)}_{klm}(\eta_0)f(\eta_0-r)D_1(\eta_0-r)
  k^{-2}{\cal Y}_{klm}(r,\bft){}^{|r},
\label{expB}
\end{eqnarray}
where we defined
\begin{eqnarray}
  f(\eta)
  ={{\cal Z}(\eta)\over H_0}
  {\partial D_1(\eta) \over \partial \eta}{1\over D_1(\eta)}
  ={ d \ln D_1(\eta) \over d \ln a(\eta)}~,
\end{eqnarray}
and we used equations (\ref{Friedman}), (\ref{deltar}), 
(\ref{ddb}), (\ref{ddD}), and (\ref{vfinal}).
Substituting equations (\ref{expA}) and (\ref{expB}) 
into equation (\ref{AC7}), we obtain
\begin{eqnarray}
  &&{\cal W}_{\rm s}(R)=
  {1\over V^{\rm LC}}
  \int{d\Omega_{\hat \bfR}\over 4\pi} 
  \int dr_1 r_1^2 \int d\Omega_{\bft_1} 
  \int dr_2 r_2^2 \int d\Omega_{\bft_2}
\nonumber
\\
  &&
 \hspace{1cm}\times
   \n0LC(r_1) \n0LC(r_2) D_1(\eta_0-r_1)D_1(\eta_0-r_2)  
\nonumber
\\
  &&
 \hspace{1cm}\times
  \int dk_1 \sum_{l_1,m_1}   \int dk_2 \sum_{l_2,m_2}
  \Bigl< \delta_{k_1l_1m_1}^{\rm (c)} (\eta_0)
  \delta_{k_2l_2m_2}^{\rm (c)*}(\eta_0)\Bigr> 
  Y_{l_1m_1}(\Omega_{\bft_1})
  Y_{l_2m_2}^*(\Omega_{\bft_2})
\nonumber
\\
  &&
 \hspace{1cm}\times\prod_{i=1}^{2}
  \biggl[b(k_i;\eta_0-r_i) X_{k_i}^{l_i}(r_i)+
  f(\eta_0-r_i){k_i^{-2}}X_{k_i}^{l_i}(r_i)^{|r_i}
  {\partial\over \partial r_i}
  \biggr]  \delta^{(3)}(\bfx_1-\bfx_2-\bfR).
\nonumber
\\
\label{C8}
\end{eqnarray}
In addition, we use the relations 
\begin{eqnarray}
  &&\delta^{(3)}(\bfx_1-\bfx_2-\bfR)={1\over (2\pi)^3}
  \int d^3\bfk ~e^{-i\bfk\cdot(\bfx_1-\bfx_2-\bfR)},
\\
  &&e^{-i\bfk\cdot\bfx}= 4\pi \sum_{l} \sum_{m=-l}^{l}
  (-i)^l j_{l}(k|\bfx|) Y_{lm}(\Omega_{\hat\bfk})
  Y_{lm}^*(\Omega_{\hat\bfx}),
\end{eqnarray}
then  equation (\ref{C8}) becomes
\begin{eqnarray}
  &&{\cal W}_{\rm s}(R)=
  {1\over V^{\rm LC}}
  \int{d\Omega_{\hat \bfR}\over 4\pi} 
  \int dr_1 r_1^2 \int d\Omega_{\bft_1} 
  \int dr_2 r_2^2 \int d\Omega_{\bft_2}
\nonumber
\\
  &&
 \hspace{1cm}\times
   \n0LC(r_1) \n0LC(r_2) D_1(\eta_0-r_1)D_1(\eta_0-r_2)  
\nonumber
\\
  &&
 \hspace{1cm}\times
  \int dk_1 \sum_{l_1,m_1}   \int dk_2 \sum_{l_2,m_2} 
  \Bigl< \delta_{k_1l_1m_1}^{\rm (c)} (\eta_0)
  \delta_{k_2l_2m_2}^{\rm (c)*}(\eta_0)\Bigr> 
  Y_{l_1m_1}(\Omega_{\bft_1})
  Y_{l_2m_2}^*(\Omega_{\bft_2})
\nonumber
\\
  &&
 \hspace{1cm}\times\prod_{i=1}^{2}
  \biggl[b(k_i;\eta_0-r_i) X_{k_i}^{l_i}(r_i)+
  f(\eta_0-r_i){k_i^{-2}}X_{k_i}^{l_i}(r_i)^{|r_i}
  {\partial\over \partial r_i}
  \biggr]
\nonumber
\\
  &&\hspace{1cm}\times
  {1\over (2\pi)^3}\int d^3\bfk
  ~4\pi \sum_{L_1M_1} (-i)^{L_1} j_{L_1}(kr_1) Y_{L_1M_1}(\Omega_{\hat\bfk})
  Y_{L_1M_1}^*(\Omega_{\bft_1}) 
\nonumber
\\
  &&\hspace{3.3cm}
  \times
  4\pi \sum_{L_2M_2} (i)^{L_2} j_{L_2}(kr_2) Y_{L_2M_2}^*(\Omega_{\hat\bfk})
  Y_{L_2M_2}(\Omega_{\bft_2}) 
\nonumber
\\
  &&\hspace{3.3cm}
  \times
  4\pi \sum_{L_3M_3} (i)^{L_3} j_{L_3}(kR) Y_{L_3M_3}^*(\Omega_{\hat\bfk})
  Y_{L_3M_3}(\Omega_{\hat {\bf R}})~, 
\label{C61}
\end{eqnarray}
where $k=|\bfk|$ and $\hat \bfk=\bfk/|\bfk|$.
Integrating over $\Omega_{\bft_1}$, $\Omega_{\bft_2}$ and
$\Omega_{{\hat {\bf R}}}$ yields
\begin{eqnarray}
  &&{\cal W}_{\rm s}(R)=
  {1\over V^{\rm LC}}
  \int dr_1 r_1^2 \int dr_2 r_2^2 
  \n0LC(r_1) \n0LC(r_2) D_1(\eta_0-r_1)D_1(\eta_0-r_2)  
\nonumber
\\
  &&
 \hspace{1cm}\times
  \int dk_1 \sum_{l_1,m_1}   \int dk_2 \sum_{l_2,m_2} 
  \Bigl< \delta_{k_1l_1m_1}^{\rm (c)} (\eta_0)
  \delta_{k_2l_2m_2}^{\rm (c)*}(\eta_0)\Bigr> 
\nonumber
\\
  &&
 \hspace{1cm}\times\prod_{i=1}^{2}
  \biggl[b(k_i;\eta_0-r_i) X_{k_i}^{l_i}(r_i)+
  f(\eta_0-r_i){k_i^{-2}}X_{k_i}^{l_i}(r_i)^{|r_i}
  {\partial\over \partial r_i}
  \biggr]
\nonumber
\\
  &&\hspace{1cm}\times{(4\pi)^2 \over (2\pi)^3}\int d^3\bfk
  (-i)^{l_1-l_2} j_{l_1}(kr_1)j_{l_2}(kr_2)j_{0}(kR) 
  Y_{l_1m_1}(\Omega_{\hat\bfk}) Y_{l_2m_2}^*(\Omega_{\hat\bfk}).
\label{C63}
\end{eqnarray}
And the further integration over $\Omega_{\hat {\bfk}}$ gives
\begin{eqnarray}
  &&{\cal W}_{\rm s}(R)=
  {1\over V^{\rm LC}}
  \int dr_1 r_1^2 \int dr_2 r_2^2 
  \n0LC(r_1) \n0LC(r_2) D_1(\eta_0-r_1)D_1(\eta_0-r_2)  
\nonumber
\\
  &&
 \hspace{1cm}\times
  \int dk_1 \int dk_2 \sum_{l}(2l+1)    
  P(k_1)\delta(k_1-k_2)
\nonumber
\\
  &&
 \hspace{1cm}\times\prod_{i=1}^{2}
  \biggl[b(k_i;\eta_0-r_i) X_{k_i}^{l}(r_i)+
  f(\eta_0-r_i){k_i^{-2}}X_{k_i}^{l}(r_i)^{|r_i}
  {\partial\over \partial r_i}
  \biggr]
\nonumber
\\
  &&\hspace{1cm}\times{(4\pi)^2 \over (2\pi)^3}\int dk k^2
  j_{l}(kr_1)j_{l}(kr_2)j_{0}(kR)~, 
\label{C65}
\end{eqnarray}
where we used the relation of the gaussian random field 
in the linear theory:
\begin{equation}
  \Bigl< \delta_{k_1l_1m_1}^{\rm (c)} (\eta_0)
  \delta_{k_2l_2m_2}^{\rm (c)*}(\eta_0)\Bigr> 
  =P(k_1)\delta(k_1-k_2)\delta_{l_1l_2}\delta_{m_1m_2},
\label{eq:q12}
\end{equation}
and $\delta_{l_1l_2}$ and $\delta_{m_1m_2}$ are the Kronecker's
delta.
Integration by parts yields:
\begin{eqnarray}
  &&{\cal W}_{\rm s}(R)=
  {1\over V^{\rm LC}}
  \int dr_1 r_1^2 \int dr_2 r_2^2 
  \n0LC(r_1) \n0LC(r_2) D_1(\eta_0-r_1)D_1(\eta_0-r_2)  
\nonumber
\\
  &&
 \hspace{1cm}\times
  \int dk_1 k_1^2P(k_1)  \sum_{l}(2l+1)    
  {4 \over\pi^2}\int dk k^2
  j_{l}(kr_1)j_{l}(kr_2)j_{0}(kR). 
\nonumber
\\
  &&
 \hspace{1cm}\times\prod_{i=1}^{2}
  \biggl[
  \biggl(b(k_1;\eta_0-r_i) - k_1^{-2}\calS_{r_i}\biggr)j_{l}(k_1r_i)
  \biggl]~,
\label{C677}
\end{eqnarray}
where we used (\ref{Xkl}), and the operator $\calS_r$ is 
defined by equation (\ref{defcalS}). Here the boundary terms are omitted
since we can show that the boundary terms are the higher order terms of
${\cal O}(R/\rmax)$ or ${\cal O}(R/\rmin)$, which are negligible 
in the case $R\ll\rmin$ and $R\ll\rmax$.

Noting the relation (e.g., Magnus et~al. 1966):
\begin{eqnarray}
  \int dk k^2 j_{l}(kr_1)j_{l}(kr_2)j_{0}(kR)
  = \left\{
      \begin{array}{ll}
        {\pi\over 4r_1r_2 R}
  P_l\biggl({r_1^2+r_2^2-R^2\over 2r_1r_2}\biggr) &
        \mbox{($|r_1-r_2|< R< r_1+r_2$)}, \\ 
        0 &
        \mbox{($R< |r_1-r_2|$, $R>r_1+r_2 $)}, 
      \end{array}
   \right. 
\end{eqnarray}
we find
\begin{eqnarray}
  &&{\cal W}_{\rm s}(R)=
  {1\over V^{\rm LC}}{1\over \pi R}
  \int\int_{\cal S} dr_1 dr_2 r_1 r_2 
  \n0LC(r_1) \n0LC(r_2) D_1(\eta_0-r_1)D_1(\eta_0-r_2)  
\nonumber
\\
  &&
 \hspace{1cm}\times
  \int dk_1 k_1^2P(k_1)  \sum_{l}(2l+1)    
  P_l\biggl({r_1^2+r_2^2-R^2\over 2r_1r_2}\biggr)
\nonumber
\\
  &&
 \hspace{1cm}\times\prod_{i=1}^{2}
  \biggl[
  \biggl(b(k_1;\eta_0-r_i) - k_1^{-2}\calS_{r_{i}}\biggr)j_{l}(k_1r_i)
  \biggr]~,
\label{C67}
\end{eqnarray}
where $\cal S$ denotes the region $|r_1-r_2|\leq R\leq r_1+r_2$.

%
By using the additional theorem for the spherical Bessel function:
\begin{equation}
  \sum_{l} (2l+1) P_l(\cos \theta) j_l(kr_1) j_l(kr_2)
  =j_0\Bigl(k\sqrt{r_1^2+r_2^2-2r_1r_2\cos\theta}\Bigr),
\end{equation}
we can write
\begin{eqnarray}
  &&{\cal W}_{\rm s}(R)=
  {1\over V^{\rm LC}}{1\over \pi R}
  \int\int_{\cal S} dr_1 dr_2 r_1 r_2 
  \n0LC(r_1) \n0LC(r_2) D_1(\eta_0-r_1)D_1(\eta_0-r_2)  
\nonumber
\\
  &&
 \hspace{1cm}\times  \int dk k^2 P(k)
  \prod_{i=1}^{2}
  \biggl[b(k;\eta_0-r_i) - k^{-2}\calS_{r_i}\biggr]
  j_0\Bigl(k\sqrt{r_1^2+r_2^2-2r_1r_2\cos\theta}\Bigr)~,
\label{C688}
\end{eqnarray}
where $\cos\theta$ is replaced by
$  \cos\theta=(r_1{}^2+r_2{}^2-R^2)/2r_1r_2$
after operating the differentiations by $r_1$ and $r_2$.

Introducing the notation $z=\sqrt{r_1^2+r_2^2-2r_1r_2\cos\theta}$, 
we can show the formulas:
\begin{eqnarray}
  &&k^{-2}{\partial^2\over \partial r^2_1}
  j_0(k z)={j_2(kz)\over z^2}(r_1-r_2\cos\theta)^2-{j_1(kz)\over kz},
\\
  &&k^{-2}{\partial^2\over \partial r^2_2}
  j_0(k z)={j_2(kz)\over z^2}(r_2-r_1\cos\theta)^2-{j_1(kz)\over kz},
\end{eqnarray}
and
\begin{eqnarray}
  &&k^{-4}{\partial^2\over \partial r_1^2}{\partial^2\over \partial r_2^2}
  j_0(k z)={j_4(kz)\over z^4}(r_1-r_2\cos\theta)^2(r_2-r_1\cos\theta)^2
  -{j_3(kz)\over kz^3}\biggl\{(r_1-r_2\cos\theta)^2
\nonumber
\\
  &&\hspace{3cm}-4\cos\theta
  (r_1-r_2\cos\theta)(r_2-r_1\cos\theta)+(r_2-r_1\cos\theta)^2\biggr\}
\nonumber
\\
  &&\hspace{3cm}+{j_2(kz)\over (kz)^2}(2\cos^2\theta+1)~.
\end{eqnarray}

Using these formulas and omitting the second term
in the derivative, i.e., 
$\calS_r\simeq f(\eta_0-r)\partial^2/\partial r^2$, 
we have 
\begin{eqnarray}
  &&{\cal W}_{\rm s}(R)\simeq{1\over V^{\rm LC}}{1\over \pi R}
  \int\int_{\cal S} dr_1 dr_2 r_1 r_2 
  \n0LC(r_1) \n0LC(r_2)
\nonumber
\\
  &&
 \hspace{1cm}\times\int dk k^2 P(k) 
  \prod_{j=1}^2 \Bigl[b(k;\eta_0-r_j)D_1(\eta_0-r_j)\Bigr]
\nonumber
\\
  &&
 \hspace{1cm}\times\Bigl[j_0(kR)+
  \beta(k;\eta_0-r_2)I(R;r_1,r_2)+
  \beta(k;\eta_0-r_1)I(R;r_2,r_1)
\nonumber
\\
  &&
 \hspace{4cm}+\beta(k;\eta_0-r_1)\beta(k;\eta_0-r_2)J(R;r_1,r_2)
  \Bigr]~,
\label{WRRA}
\end{eqnarray}
where $\beta(k;\eta)$ is defined by equation (\ref{defbeta}),
and $I(R;r_1,r_2)$ and $J(R;r_1,r_2)$ are defined by 
\begin{equation}
  I(R;r_1,r_2)={j_1(kR)\over kR}-{j_2(kR)\over R^2}
  \biggl\{{R^2+r_2^2-r_1^2\over 2r_2}\biggr\}^2~,
\label{defI}
\end{equation}
and
\begin{eqnarray}
  &&J(R;r_1,r_2)={j_2(kR)\over (kR)^2}
  \biggl[2\biggl\{{r_1^2+r_2^2-R^2\over 2r_1r_2}\biggr\}^2+1\biggr]
\nonumber
\\
  &&
 \hspace{2.2cm}
  +{j_4(kR)\over R^4}\biggl\{{R^2+r_1^2-r_2^2\over 2r_1}\biggr\}^2
                     \biggl\{{R^2+r_2^2-r_1^2\over 2r_2}\biggr\}^2
\nonumber
\\
  &&
 \hspace{2.2cm}-{j_3(kR)\over kR^3}\biggl[
  \biggl\{{R^2+r_1^2-r_2^2\over 2r_1}\biggr\}^2+
  \biggl\{{R^2+r_2^2-r_1^2\over 2r_2}\biggr\}^2
\nonumber
\\
  &&
 \hspace{4cm}  -
  {R^2+r_1^2-r_2^2\over r_1}{R^2+r_2^2-r_1^2\over r_2}
  {r_1^2+r_2^2-R^2\over 2r_1r_2}\biggr],
\label{defJ}
\end{eqnarray}
respectively. 

Since we are generally interested in the case 
of $R\ll\rmax$, we can use the approximation:
\begin{eqnarray}
  \int\int_{\cal S} dr_1 dr_2\simeq 
  \int_\rmin^\rmax dr_1 \int_{-R}^{R} dx ,
\end{eqnarray}
where we introduced $x=r_2- r_1$. By expanding $I(R;r_1,r_2)$
and $J(R;r_1,r_2)$ in terms of $x$, we have
\begin{eqnarray}
  &&I(R;r_1,r_2)={j_1(kR)\over kR}-{j_2(kR)\over R^2}x^2
\\
  &&J(R;r_1,r_2)=3{j_2(kR)\over (kR)^2}
  -6{j_3(kR)\over kR^3} x^2+{j_4(kR)\over R^4} x^4
\end{eqnarray}
Integration by  $x$ leads to the final expression: 
\begin{eqnarray}
  &&{\cal W}_{\rm s}(R)\simeq
  {4\pi\over V^{\rm LC }}
  \int_\rmin^\rmax dr r^2 \n0LC(r)^2 
  {1\over 2\pi^2}\int  k^2 dk  P(k)  
  \Bigl[b(k;\eta_0-r)D_1(\eta_0-r)\Bigr]^2
\nonumber
\\
  && \hspace{1.5cm}\times
  \biggl[1+{2\over 3}\beta(\eta_0-r)+{1\over5}\beta(\eta_0-r)^2
  \biggr]j_0(kR)~ .
\label{WRA}
\end{eqnarray}

\bigskip
\clearpage

\parskip2pt
\centerline{\bf REFERENCES}
\bigskip

\def\apjpap#1;#2;#3;#4; {\pp#1, {#2}, {#3}, #4}
\def\apjbook#1;#2;#3;#4; {\pp#1, {#2} (#3: #4)}
\def\apjppt#1;#2; {\pp#1, #2.}
\def\apjproc#1;#2;#3;#4;#5;#6; {\pp#1, {#2} #3, (#4: #5), #6}
\apjpap Bardeen,~J.M., Bond,~J.R, Kaiser,~N., \& Szalay,~A.S. 1986;
 ApJ;304;15;
\apjppt Boyle, B.J., Croom, S.M., Smith, R.J., Shanks, T., Miller L., 
 \& Loaring, N. 1998;Phil.Trans.R.Soc.Lond.A, in press (astro-ph/9805140);
\apjpap Carrera, F.J. et~al. 1998;MNRAS;299;229;
\apjpap Cress, C.M., Helfand, D.J., Becker, R.H., Gregg, M.D., 
  \& White, R.L. 1996;ApJ;473;7;
\apjpap Croom, S.M. \& Shanks, T.1996;MNRAS;281;893;
\apjpap Davis, M. \& Peebles, P.J.E. 1983;ApJ;267;465;
\apjppt Dekel, A. \& Lahav, O. 1998;ApJ, submitted (astro-ph/9806193);
\apjpap de Laix, A. A. \& Starkman, G. D. 1998;MNRAS;299;977;
\apjpap Fry, J. N. 1996;ApJ;461;L65;
\apjpap Giavalisco, M., Steidel, C.C., Adelberger, K.L., 
  Dickinson, M.,  Pettini, M., \& Kellogg, M. 1998;ApJ;503;543;
\apjppt Hamilton, A.J.S. 1997; to appear in the Proceedings of
  Ringberg Workshop on Large-Scale Structure, edited by Hamilton, D.
  (astro-ph/9708102);
\apjpap Hamilton, A.J.S. \&  Culhane, M. 1996;MNRAS;278;73; 
\apjpap Heavens, A.F. \& Taylor, A.N. 1995;MNRAS;275;483; 
\apjpap Jing, Y.P.,  \& Suto, Y. 1998;ApJ;494;L5;
\apjpap Kaiser, N. 1987;MNRAS;227;1;
\apjpap Kitayama, T. \& Suto, Y. 1997;ApJ;490;557;
\apjpap Kodama, H. \& Sasaki, M. 1984;Prog. Theor. Phys. Supp.;78;1;
\apjppt Magliocchetti. M.,  Maddox, S.J.,  Lahav, O.,  \& Wall, J.V. 1998;
  MNRAS, submitted (astro-ph/9806342);
\apjpap Matarrese, S., Coles, P., Lucchin, F., \& Moscardini, L. 1997;
 MNRAS;286;115;
\apjpap Matsubara, T. \& Suto, Y. 1996;ApJ;470;L1;
\apjpap Matsubara, T. , Suto, Y., \& Szapudi,I. 1997;ApJ;491;L1;
\apjpap Mo, H.J. \& White, S.D.M. 1996;MNRAS;282;347;
\apjpap Nakamura, T.T., Matsubara, T., \& Suto, Y. 1998;ApJ;494;13;
\apjppt Peacock, J.A. 1998;Phil.Trans.R.Soc.Lond.A, 
  in press (astro-ph/9805208);
\apjpap Steidel, C.C., Adelberger, K.L., Dickinson, M., Giavalisco,
  M., Pettini, M., \& Kellogg, M. 1998;ApJ;492;428;
\apjpap Sugiyama,~N. 1995;ApJS;100;281;
\apjppt Suto, Y., Magira, H., Jing, Y.P., Matsubara, T., \& Yamamoto, K. 1999;
  Prog. of Theor. Phys. Suppl., in press (astro-ph/9901179);
\apjpap Szalay, A.S., Matsubara, T., \& Landy, S.D. 1998;ApJ;498;L1;
\apjppt Taruya, A., Koyama, K., \& Soda, J. 1998;ApJ, in press 
(astro-ph/9807005);
\apjpap Tegmark, M. \& Peebles, P.J.E. 1998;ApJ;500;L79;
\apjppt Yamamoto, K. \& Suto, Y. 1999;ApJ, in press (Paper~I);
\clearpage

\newpage
\begin{figure}
\begin{center}
    \leavevmode\psfig{file=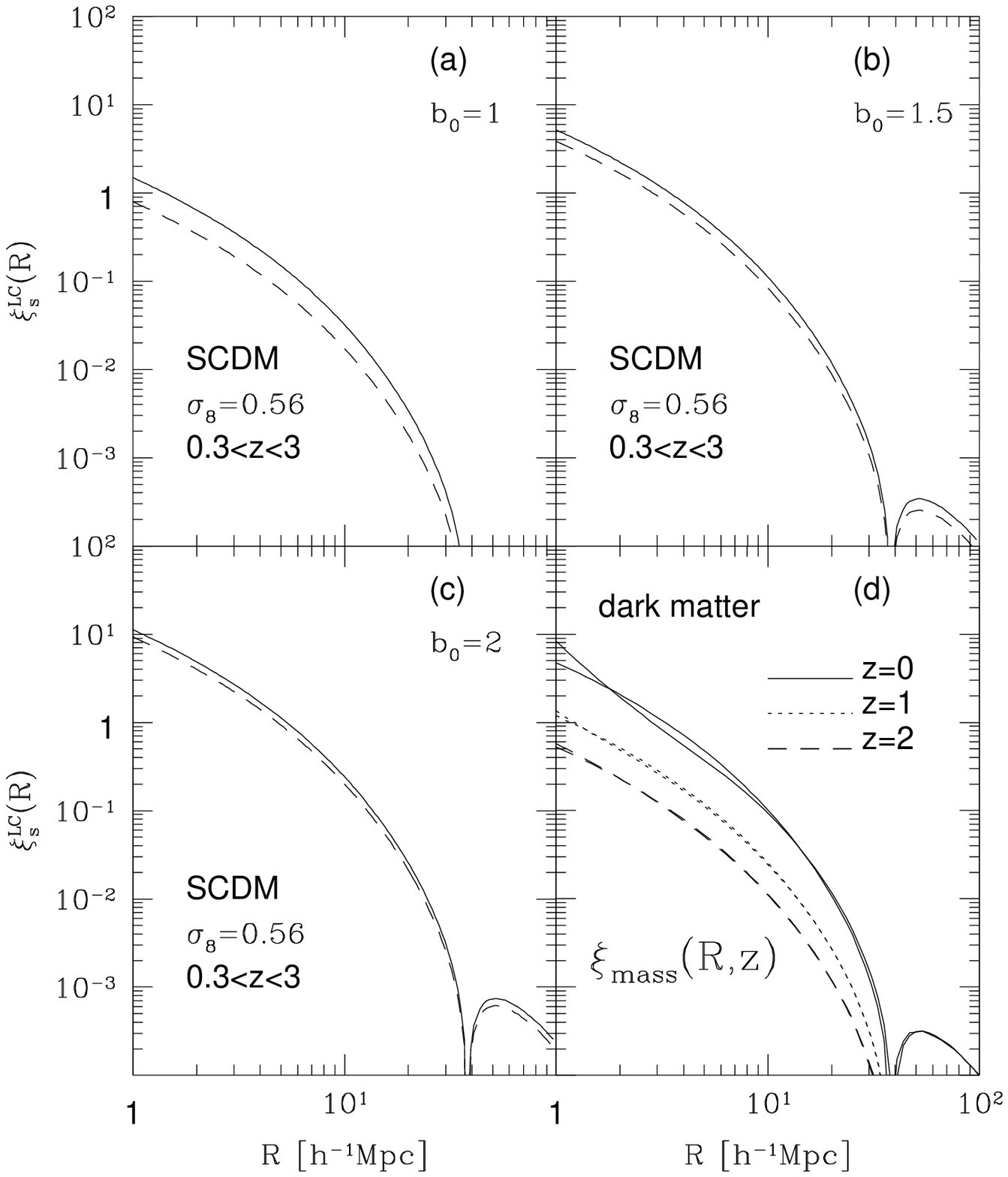,width=12cm}
\end{center}
\figcaption
{ Absolute values of the two-point correlation function for 
QSO on a light-cone and the linear and
nonlinear mass two-point correlation functions
in the standard CDM model, where we adopted 
$\Omega_0=1,~\Omega_\Lambda=0,~h=0.5,~$ and the CDM 
density power spectrum normalized as $\sigma_8=0.56$
(Kitayama \& Suto 1997).
The parameter $b_0$ for the bias model are adopted as
(a)$b_0=1$,~(b)$b_0=1.5$,~(c)$b_0=2$.
Here we assumed that the sources are distributed in the
range $0.3\leq z\leq 3$ with a constant number density. 
In each panel (a)-(c) the solid line shows our $\xis^{\rm LC}(R)$, 
the dashed line does the case when the redshift-space 
distortion is neglected by setting $\beta=0$, 
(d) linear (lower curves) and nonlinear (upper curves) mass
correlation functions by Peacock \& Dodds (1996) defined on constant-time 
hypersurfaces $z=0,~1$ and $2$. 
\label{fig:scdm}}
\end{figure}
\clearpage
\begin{figure}
\begin{center}
    \leavevmode\psfig{file=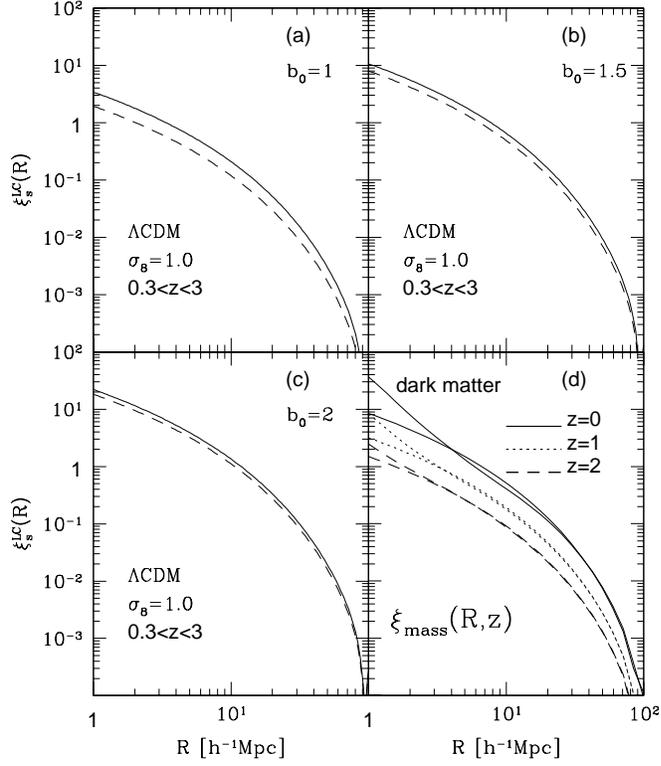,width=12cm}
\end{center}
\figcaption
{ Same as Fig.~1 but for the $\Lambda$CDM model, in which 
we adopted $\Omega_0=0.3,~\Omega_\Lambda=0.7,~h=0.7,~\sigma_8=1.0$.
\label{fig:lcdm}}
\end{figure}
\end{appendix}
\end{document}